\documentclass{aa}
\usepackage{graphicx}
\begin{document}
\title{Kinematic frames and ``active longitudes'': does the Sun have a face?}
   \author{J. Pelt
          \inst{1}
          \and
          J.M. Brooke
          \inst{2,3}
          \and
          M.J. Korpi
          \inst{4}
          \and
          I. Tuominen
          \inst{4} 
          }
   \offprints{J. Pelt}
   \authorrunning{J. Pelt et al}
   \titlerunning{Kinematic frames and ``active longitudes''} 
   \institute{
             Tartu Observatory, 61602 T\~{o}ravere, Estonia
             \and
             Manchester Computing, University of Manchester, Oxford Road, 
             Manchester, M13 9PL, UK 
             \and
             School of Computer Science, University of Manchester, Oxford Road, 
             Manchester, M13 9PL, UK  
             \and
             Observatory, PO Box 14, FI-00014 University of Helsinki, Finland           
             }
             
             \date{Received ---; accepted ---} 
       
\abstract{It has recently been claimed that analysis of Greenwich sunspot data
   over 120 years reveals that sunspot activity clusters around two longitudes separated
   by 180\degr (``active longitudes'') with clearly defined differential
   rotation during activity cycles. In previous work we demonstrated that
   such effects can be observed in synthetic data without such features,
   as an artefact of the method of analysis.}
  {In the present work we extend this critical examination of methodology to
   the actual Greenwich sunspot data and also consider newly proposed methods of analysis
   claiming to confirm the original identification of active longitudes.}
  {We performed fits of different kinematic frames onto the actual sunspot data.
   Firstly, a cell-counting statistic was used to analyse a comoving system of frames
   and show that such frames extract useful information from the data. 
   Secondly, to check the
   claim of century-scale persistent active longitudes in a contramoving frame
   system, we made a comprehensive exploration of parameter space following the 
   original methodology as closely as possible.}
  {Our analysis revealed that values obtained for
   the parameters of differential rotation are not stable across different methods of
   analysis proposed to track persistent active longitudes. 
   Also, despite a very thorough search in parameter space, we were unable
   to reproduce results claiming to reveal the century-persistent active longitudes. 
   Previous parameter space exploration has been restricted to frames whose latitudinal 
   profile is opposite to solar surface differential rotation. 
   Relaxing this restriction we found that the
   highest values of nonaxisymmetry occur for frames comoving with the solar surface flow. 
   Further analysis indicates that even these solutions
   are the result of purely statistical fluctuations.}
  {We can therefore say that strong and well substantiated evidence for 
   an essential and century-scale persistent nonaxisymmetry in the sunspot
   distribution does not exist.}

\maketitle
\section{Introduction}\label{Section:introduction}
\subsection{Nonaxisymmetry in late-type stars}
It is known that solar active regions may exist at the same
position (such as recurrent sunspot groups) during several solar
rotations, a new active region appearing at another position (e.g.
Tuominen \& Virtanen~\cite{TV87}, Pulkkinen \&
Tuominen~\cite{PT98}). The sunspots reflect intense concentrations of
magnetic field and thus occur only at certain discrete locations on
the solar surface. For related discussion on the observed magnetogram
recordings of the full-disk field see e.g. Stenflo~\cite{Stenflo91}.
Consequently, different activity indicators tend to be statistically
correlated for a long time spans and a statistical analysis of
the corresponding phase distributions must take this into account. 

 The theory of the global magnetic field
(mean-field dynamo theory) has been developed over the last 50 years,
starting from a linear theory (Parker~\cite{Parker55}, Steenbeck \&
Krause~\cite{SK69}, for a detailed review see Krause and
R\"adler~\cite{KR80}), followed by a nonlinear one, where the
stability of the solutions determines the symmetry properties of the
field (Krause \& Meinel~\cite{KM88}, Brandenburg et
al.~\cite{Betal89}), and including a nonaxisymmetric solution as a
possibility in the case of rapid rotation (e.g R\"adler et
al.~\cite{Retal90}, Moss, Tuominen \& Brandenburg~\cite{MTB91}). More
recent models including hydrodynamics show that when the rotation
velocity increases, the field geometry changes from antisymmetric with
respect to the equatorial plane and axisymmmetric with respect to the
longitude, to nonaxisymmetric with respect to the longitude
(e.g. Tuominen et al.~\cite{TBKR99}, Tuominen, Berdyugina \&
Korpi~\cite{Tuominen2002}).

Recently it has been claimed that long-term, i.e. persistent for at least 120
years, nonaxisymmetric structures can be detected in the solar sunspot data
using new methods of analysis that take into account the possibility that such
structures follow the differential surface rotation. This is an important
claim and in the present paper we investigate it in detail because the existence of
persistent ``active longitudes'' (nonaxisymmetric structures) in old
slowly rotating late-type stars, such as the Sun, has important
theoretical implications.

\subsection{Nonaxisymmetry in solar data}
Berdyugina \& Usoskin~\cite{BU} (hence
BU) analysed the longitudinal distribution of regions of sunspot formation in the 
Greenwich sunspot data for a period of 120 years. By applying phase
  corrections, corresponding to the differential rotation of regions of activity
  formation, they claimed to identify persistent
(during 120 years) active longitude belts which are separated by $180\degr$
and which drift smoothly along longitudes. The essence of their method is
to apply a longitude shift correction for each Carrington rotation to
represent a different rotation velocity for the active longitudes as the 
sunspot
activity wave moves from high to low latitudes in each solar cycle. 
As a response to the first paper (BU)
Pelt, Tuominen \& Brooke ~\cite{PTB} (PTB)
demonstrated that large part (or all) of the persistent smooth migration of
the active longitudes can be the result of a
particular method of data treatment 
in BU. Essentially it was shown that diagrams, similar to
these used as a proof in BU, can be obtained even from random data. The
criticisms of the paper were not directed at the observation that there can
be active longitudes which persist over time, but towards the claim that such
longitudes maintain consistent 
migration on a century time-scale.
 
In recent papers Usoskin, Berdyugina \& Poutanen~\cite{UBP} (UBP) and  
Berdyugina et al.~\cite{BMSU} (BMSU) responded to this critique with
the claim that they can 
confirm the original BU results using
much simpler constructions which do not use the data processing methods
criticized in PTB. In particular they consider that the new analysis confirms
that the active longitudes maintain consistent smooth migration
throughout the 120 year period of the sunspot data analysed. The fact that
this result was produced by two different methods of analysis of the data is
considered to confirm the reality of the phenomenon.

As will be discussed in length below, the migration
models given in the first paper (BU) significantly differ from those
obtained as a result of the analysis in UBP and represented in the
data analysis part of BMSU. This is most readily seen in the
cumulative frame kinematics: the BU frame makes about 28 extra
rotations compared to the Carrington frame, contrasted with the UBP
values of about $\approx 11$ rotations for the Northern hemisphere and
about $\approx 14$ for the Southern hemisphere.  The particular key
feature here is that in BU the Northern and Southern frames keep
coherently in phase whereas in UBP the Southern frame lags with
respect to the Northern one.
This indicates the need for a
very thorough analysis of the methods of analysis used in constructing the 
differentially rotating frames considered to ``freeze'' the longitudes of
these nonaxisymmetic features.

This paper is structured as follows: in section~\ref{Section:data} we describe the data and the 
background to our analysis. In section~\ref{Section:cellcounting} we demonstrate that kinematic frames 
which follow the rotation of sunpots exist and can clarify the basic
properties of the longitudinal distribution of sunspots. 
In section~\ref{Section:method} we present an examination of parameter 
space using the methods used in UBP but increasing the resolution and the range of the 
parameter space search. In section~\ref{Section:discussion} we discuss these results and introduce 
clarificatory statistical computations to check whether bimodal distributions
could be statistical fluctuations. Finally section~\ref{Section:conclusions} presents our conclusions.

\section{Data and methods of analysis}\label{Section:data}
\subsection{Details of datasets analysed}
Our input data for statistical analysis was downloaded from the same
Science at NASA web
site\footnote{http://science.msfc.nasa.gov/ssl/pad/solar/greenwch.htm}
as in BU and UBP. The format description supplied with the data
sets  turned out to be incomplete and to unscramble the data we used
the original  document from NOAA Satellite and Information Service
Site\footnote{ftp://ftp.ngdc.noaa.gov/STP/SOLAR\_DATA/\break
SUNSPOT\_REGIONS/GREENWICH/GROUPS/format.grp}.  During the data
checking we found some outliers in the data. The record from
observations of year 1980 which referred to longitude $408.9\degr$
was rejected and for couple of  records where longitude was only
slightly more than $360\degr$ we  subtracted $360\degr$ from the
given value (assuming circularity).  The subset 
for years 1878-1996 was then singled out to be 
compatible with papers under scrutiny. In the terms of Carrington
rotations we used rotations 325-1917 for both hemispheres. For each
particular sunspot group or single spot we selected only the record
of its first occurence as was done in original papers.  It is
important to note that some sunspot group numbers in data base are 
multiply used. To count them as separate entities we used the
index value from the 21st column in the records  (this was the
piece of information missing in the first  format
description mentioned above). Finally we had two data sets: 1593 
rotations and 18680 records for the Northern hemisphere and 1593
rotations  and 17966 records for the  Southern hemisphere; the
compiled data sets are available on the
web\footnote{http://www.aai.ee/\~{}pelt/soft.htm}.  In the original
papers ``about 1600'' rotations and ``about 40000'' records were
used. 

To simplify the discussions below we will often renumber
the original Carrington rotation numbers in our data set to start
from 1 (hence  $1,\dots,1593$). Furthermore, in all the plots and
formulae  we adopt the basic phase interval $0\degr-360\degr$. As a
result, some formulae can differ from those of the original
papers. However, the original forms can easily be recovered by using
an appropriate phase shift.

\subsection{Construction of kinematic frames for analysis}\label{Subsection:comoving}
Two different physical scenarios can be considered when thinking of
how to construct a framesystem. If one is to construct a frame to
correct for the solar surface differential rotation, it is natural to
construct a frame that is moving with this flow. To investigate the
situation where a rigidly rotating nonaxisymmetric structure anchored
to the solar core would "illuminate" the solar activity belt
influenced by differential rotation (the stroboscopic explanation
discussed by BMSU), a more natural starting point would be to
investigate frames which move against the solar surface differential rotation
pattern. As both situations could occur in the Sun, neither of them
should be excluded from the analysis.

Let us now divide the full set of sunspot records into a set of
subintervals of length 27.2753 days, the Carrington
rotation, as proposed originally by BU. For each particular rotation
$i, i=1,\dots,N$ (where $N$ is the total number of rotations in
the set) the local phase shift $\Omega_i$ is defined (in degrees per
day). It can be constant or depend on a mean latitude of sunspot
formation and free parameter $B$ as in Eq.~\ref{equation2a}.  If the
frame is to follow the differential rotation pattern then, over
the full time span, the shifts accumulate  according to the formula
\begin{equation}
\Lambda_i=\Lambda_0+T_C\sum_{j=1}^i (\Omega_j-\Omega_C),
\label{equation1.1}
\end{equation}
i.e. the frame will lag if the sunspots rotate slower and
will be advanced for more rapid rotation. Here $T_C=25.38$,
$\Omega_C=360\degr/T_C$ and $\Lambda_0$ is  a general phase shift to be
computed later. Hereafter we refer to this frame system as the
{\it comoving} frame. The other possibility is to choose a system of frames
that moves in the opposite direction if compared to the surface
rotation pattern, when the shifts are defined as
\begin{equation}
\Lambda_i=\Lambda_0+T_C\sum_{j=1}^i (\Omega_C-\Omega_j).
\label{equation1.2}
\end{equation}
We will refer to this frame system as the {\it contramoving} frame.

If the differential rotation is defined as Eq.~\ref{equation2a}, then
we see that for the comoving frame, positive values of $B$ lead to
phases being pushed backwards if sunspots occur at higher latitudes
where they rotate slower. If we allow for negative values of $B$, in
the comoving frame they correspond to frames that are pushed forward
with slower rotation. The opposite sign rule applies for the contramoving
frames, positive $B$s lead to phases being pushed forward with slow
rotation and negative $B$s to a situation where the frame follows
solar-type differential rotation. The basic difference of the two
frame systems is illustrated in Fig.\ref{Figure04}; the comoving frame
phase shift curve resulting from the best solution from the
cell-counting statistics presented in Sect.~\ref{Section:cellcounting} is plotted with a thin
solid line and the original UBP Northern contramoving solution phase
shift on top of it with a thick solid line. In the former solution the
positive $B$-value causes the phases to lag behind during the early
stages of the solar cycle (sunspots at higher latitudes) whereas the
positive value in the latter solution results in phases being pushed
forward during the early stages.  

If we do not restrict the parameter $\Omega_0$ and allow for both
positive and negative values of the parameter $B$ within each frame
system, one can see that formally the two types of frames are equivalent,
and both types of physical situations of interest are detectable by
both methods. In this study we utilise both types of frames with no
restriction on the sign of $B$. We note here that in UBP and BMSU the
analysis was restricted to the contramoving formulation with positive
$B$ values only. 
It is a key point that, as we demonstrate below, although formally
equivalent, they are not statistically
equivalent. This is due both to the interaction of the frames with random
fluctations in longitudinal distributions and also to the method of analysis proposed in UBP.

To facilitate the comparison and transformation
between the framesystems, the contramoving parameters translate to
comoving frame by the following formulae
\begin{eqnarray}
B^{comov} &=& - B^{contra} \\
\Omega_0^{comov} &=& 2 \Omega_C - \Omega_0^{contra}. 
\end{eqnarray}
Here we also note that special attention is needed when contramoving
rotation laws are plotted together with comoving rotation laws
(e.g. observational data from sunspots or helioseismology). The
contramoving parameters must first be translated to comoving frame by
the formulae given above. For instance, the best frame for the
Northern hemisphere in UBP, obtained in the contramoving formulation
with the parameters $\Omega_0^{contra}$=14.33 and $B^{contra}$=3.40,
translates to comoving parameters $\Omega_0^{comov}$=14.05 and
$B^{comov}$= -3.40, i.e. slow rotation at the equator which cyclically
speeds up as a function of latitude. We have illustrated the possible
error related to treating the comoving and contramoving frames as
equals in our Fig.~\ref{Figure05}, where we have plotted the best Northern
UBP frame twice: the thin solid line represents the original
illustration of UBP where the rotation law with the contramoving
parameterization is plotted amongst comoving rotation laws without
transformation. With the solid thick line we plot the transformed UBP
rotation law in the comoving frame. Here it can be clearly seen
that comparisons of the UBP results with the solar surface
differential rotation indicators are not meaningful.

\section{Cell-counting method} \label{Section:cellcounting}

To illustrate how the kinematic frames work in practice, in this
section we construct and analyse one frame system with a statistical
method that is as simple as possible. We choose the comoving formulation, for
which the phase shift is defined as Eq.~\ref{equation1.1}.
Next we subtract the flow model from actual longitudes 
\begin{equation}
\tilde\lambda_{ki}=\lambda_{ki}-\Lambda_i-m_{ki} \times 360\degr, 
\label{phases}  
\end{equation}
and select integers $m_{ki}$ so that the results will lay in the
interval $[0,360\degr]$. As we now work in the comoving formulation, after
the subtraction we get a certain distribution of longitudes in a
frame which is at rest with respect to the differential rotation
pattern.

Our comoving frame, as any other frame model in general, depends
on a set of parameters (typically $\Omega_0,B$ and $\Lambda_0$).  For
each set of parameters we can compute corrected phases $\lambda_{ki}$
(see Eq.~\ref{phases}) and a value of a certain merit function
(criterion) which characterizes distribution of these phases
(longitudes). For compatibility we will use merit functions whose
parameter dependent {\it minima} are indicators of the desired
types of phase distributions.

To define our first merit fuction we divide all Carrington rotations
into groups of ten rotations and all longitudes (phases) into 6\degr ~
wide cells. We work with 1593 separate rotations altogether and
thus get a matrix of $60 \times 160$ cells. For each set of
free parameters (here $\Lambda_0$ and $\Omega_0$) we compute a merit
function in the following simple way: Each of the corrected phases
$\tilde\lambda_{ki}$ belongs to a certain cell.  Some of the cells
remain empty because there are no sunspots in this particular rotation
and interval of phases. As a criterion we use the ratio of occupied
cells to the total number of phases.
The rationale of this scheme is to produce a measure that is sensitive to the potential
existence of longitudes of enhanced spot formation. If the spots or spot groups
form at persistent longitudes (in a corresponding frame), they tend
to have their first observation in the cells which contain or are close to
this persistent longitude.
By fixing the cell length in number of rotations, we also fix the length of time
interval for which we assume the activity to be persistent. The width
of the cells fixes the amount of allowed phase dispersion during this
number of rotations. 

\subsection{Comoving frames with fixed angular velocity}
\begin{figure}
  \resizebox{\hsize}{!}{\includegraphics{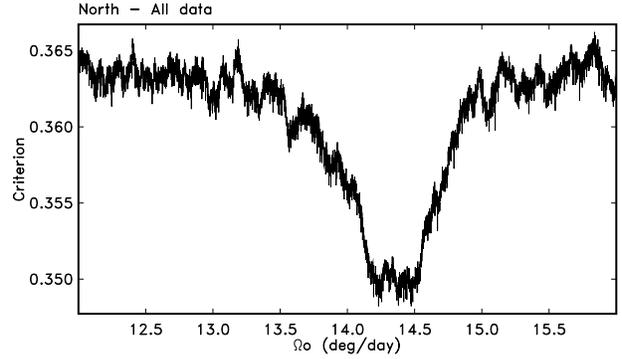}}
  \caption{The cell-counting merit function plotted against $\Omega_0$ for the comoving 
   frame with fixed angular velocity. 
  }
\label{Figure01}
\end{figure}

We now demonstrate that the formalism just described is sound and
usable. We start from the simplest system of shifts
$\Omega_i=\Omega_0$, that is, we assign a constant shift $\Omega_0$
for each Carrington rotation. We let $\Omega_0$ vary in a wide range
around the Carrington rotation velocity $\Omega_C$, and calculate the
simple cell-counting merit function described above as function of
$\Omega_0$; this is depicted in Fig.~\ref{Figure01}. The dependence on
$\Lambda_0$ is ``minimized out''. The curve is somewhat noisy, but
still demonstrates clearly how the phases start to line up for
$\Omega_0$ values larger than $\approx 14.1$ deg/day and how the
merit function value again increases from approximately
$14.6$ degrees per day. The wide depression is too noisy to single out
an optimal frame for all latitudes together.

\begin{figure}
  \resizebox{\hsize}{!}{\includegraphics{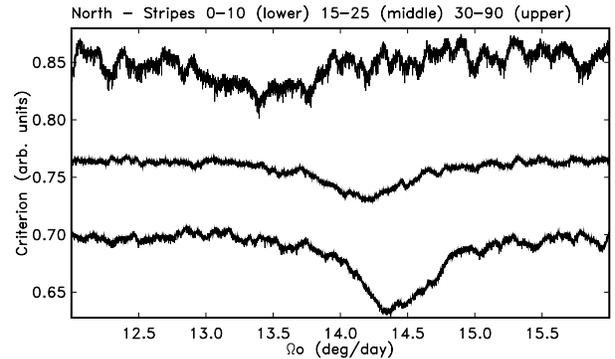}}
  \caption{Merit functions for the subsets of data in different latitude
    strips. The minima are now shifted and this is a result of differential
    rotation. 
    }
\label{Figure02}
\end{figure} 
We now generalize our method to accomodate differential rotation by
preselecting latitudinal strips and repeating the analysis on each of
these subsets separately.
We see from Fig.~\ref{Figure02} that the minima for each subset are more clearly
defined and show a shift representing the migration between latitudes with 
different rotation velosities.
The values of $\Omega_0$ for different strips can be used to build a
latitudinally dependent rotation profile. The equatorial angular velocity is
$\approx 14.37$, for middle
latitudes $\approx 14.18$ and for high latitudes $\approx 13.40$ 
degrees per day. The merit function curves are
still noisy, but they can be improved by smoothing. 
Because of the different amount of activity indicators
in different latitude strips the general level and scatter for the three curves
are different.
{By constructing this merit function we can show that
{\begin{itemize}
\item The hypothesis of comoving frames shows the 
existence of a range of preferred frames even when applied to the full 120 year  data set.
\item The scheme with phase corrections allows the computation of optimal
parameters for  different comoving frames.
\item The comoving frames in different latitude strips have stable 
rotational velocities revealed as principal merit 
function minima. 
\item The minima of the merit functions are located inside wide depressions, 
which shows that they are not minor  fluctuations manifesting 
themselves as narrow peaks.
\item The kinematics of frames corresponds to kinematics of differential rotation - 
frames at higher latitudes tend to rotate slower than frames near equator. 
\end{itemize}
In this analysis we have
 used the longitudes of formation of all sunspots (not
weighted by area) since we wish first to test the most basic hypothesis of
persistence of activity formation.
\subsection{Comoving frames with changing angular velocity}
Moving on from the basic analysis above, we now attempt to detect an optimal single frame
for all latitudes including weighting for sunspot area. Following  UBP 
we will now include  an additional term for each  Carrington 
rotation:
\begin{equation}
\Omega_i=\Omega_0-B \sin^2 \langle \psi \rangle_i,
\label{equation2a}
\end{equation}
where $\langle\psi_i\rangle_i$ is an area weighted average latitude
for $i$-th Carrington rotation and parameter $B$ measures the
amplitude of the latitude dependent part.

To get the optimal triple of frame parameters we need to compute merit 
functions for
a large grid of trials. Using  a two step procedure, 
decribed in detail below, we found  a global
minimum for  the parameter grid 
$(\Omega_0 \times \Lambda_0 \times B)=[13.5-15.0;0.001]\times[0.0-6.0;0.02]\times[1-5;1]$\footnote{Here and below
we use  a systematic notation for computation grids. 
Inside  the square brackets we
give minimum and maximum values for  the parameter  in question 
followed by  the stepsize  of computation.} 
to be located at $\Omega_0 \approx 14.416$ and $B \approx 2.32$. 
The range of $\Lambda_0$ investigated results from the fact that 
there are only six different 
starting positions for $6\degr$-wide cells. The resulting merit functions 
are depicted in
Fig.~\ref{Figure03} and  the corresponding shifts for 
the comoving frame in Fig.~\ref{Figure04}.  

\begin{figure}
   \resizebox{\hsize}{!}{\includegraphics{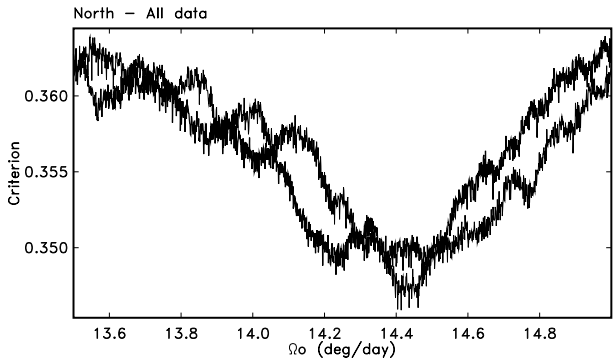}}
\caption{Merit functions computed for  the whole data set: 
            $B = 0$ (fixed  angular velocity, thick line),
            $B = 2.32$ (varying angular velocity, thin line).           
            }
\label{Figure03}
\end{figure} 

\begin{figure}
  \resizebox{\hsize}{!}{\includegraphics{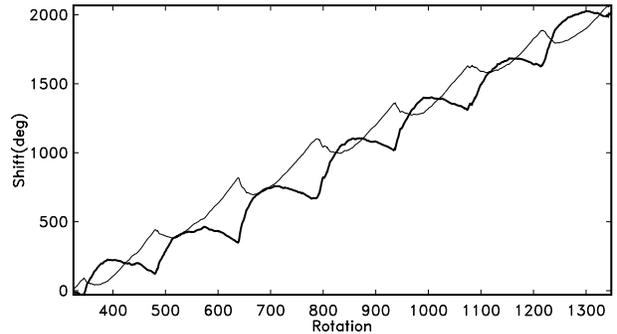}}

  \caption{Typical shift curves for comoving (thin line) and contramoving (thick line) frames.}
\label{Figure04}
\end{figure}

It is important to stress that the strongest minimum for 
the frames with changing angular velocity
is deeper  compared with that of the fixed  
angular velocity frames. Consequently it
indeed models to some extent real features in the distribution of sunspot
longitudes.

This analysis demonstrates that the formalism of
comoving frames can reproduce previous results which show that sunspots form at
longitudes which can persist for time period much longer than a Carrington rotation.
 
\section{The UBP method}\label{Section:method}
The basic claim of  BU, UBP and BMSU is that  two preferred 
longitudes separated by $180\degr$ persisting over the
last 120 years can be seen in the distribution of sunspot longitudes. 
These ``active longitudes'' migrate in any frame with 
fixed angular velocity, but form coherent and persistent longitudinal 
structures in a certain frame with 
changing angular velocity. In the previous section we have shown 
that frames with changing  angular velocity can be
used to reveal  short-term (of the order of 10 rotations) 
persistent structures.

The first difference between our simple cell-counting example method and the
method of UBP comes from 
the fact that they have chosen to use phase corrections of the contramoving
type, given by Eq.\ref{equation1.2}.
The second difference 
is the usage of sunspot areas.  Our simple merit function treats only
longitudes whereas the areas are accounted for only when computing the
mean latitudes.  In UBP the areas are first normalized and then used
as weights in the merit function. This amplifies the effect of
the spots occurring during the low activity interval of the solar
cycle. The third difference is the usage of different merit
functions.
Since in UBP the authors seek a particular
phase distribution they use a particularly crafted bimodal merit
function (see Fig.~\ref{Figure06} and Eq.~\ref{equation3}), whereas ours
seeks to minimize phase drift between cells.  

\begin{figure}
  \resizebox{\hsize}{!}{\includegraphics{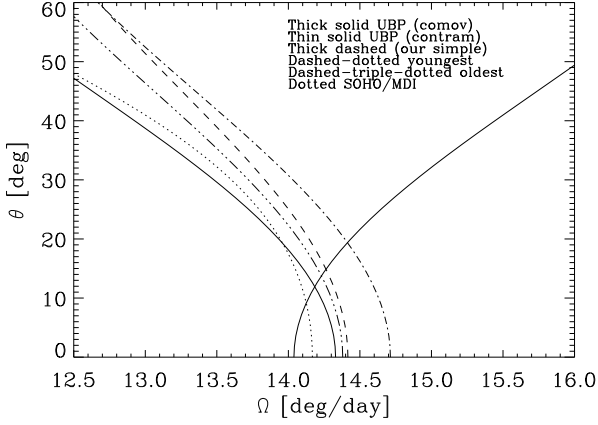}}
  \caption{Comparison of differential rotation curves obtained from helioseismology
(SOHO/MDI, Schou et al.~\cite{Schou}), for sunspots of different ages (Pulkkinen \&
Tuominen~\cite{PT98}), and for the different frame formulations investigated in
this paper.}
\label{Figure05}
\end{figure} 

Since the claims of UBP go well beyond previous analysis of
persistence of nonaxisymmetric features and lead to results which are
difficult to explain (e.g. disconnection of Northern and Southern
hemispheres) we present a very detailed and careful analysis of the
full parameter space below.
The only principal difference between our analysis in this section and the UBP investigation is
the extension of the method to negative values of $B$, i.e. we also
allow for the possibility to correct for solar-type differential
rotation. Otherwise we have followed the original implementation as carefully
as possible.
The important refinements are given in detail below. Possible
reasons why our results differ from those of UBP are given in
Sect.\ref{Section:discussion}.

\subsection{Search for the optimal frame}
For every combination of parameters $\Omega_0$, $B$ and $\Lambda_0$ 
we can build a corresponding frame. The shifted phases are 
then evaluated with the UBP method using the merit function
\begin{equation}
{\cal E}={\sum_{i} \sum_{k} A_{ik}\epsilon_{ik}^2 \over \sum_{i} \sum_{k} A_{ik}},
\label{equation3}
\end{equation}
where $\epsilon_{ik}=\min(\min(\tilde\lambda_{ik},360\degr-\tilde\lambda_{ik}),|\tilde\lambda_{ik}-180\degr|)$ measures 
the distance between the corrected phases and the 
nearest centre ($0\degr$ or $180\degr$)\footnote{In the printed version of UBP this function 
is given incorrectly. This mistake was also 
confirmed by I. Usoskin, one of the authors of UBP ~\cite{UsoskinP}}. 

\begin{figure}
  \resizebox{\hsize}{!}{\includegraphics{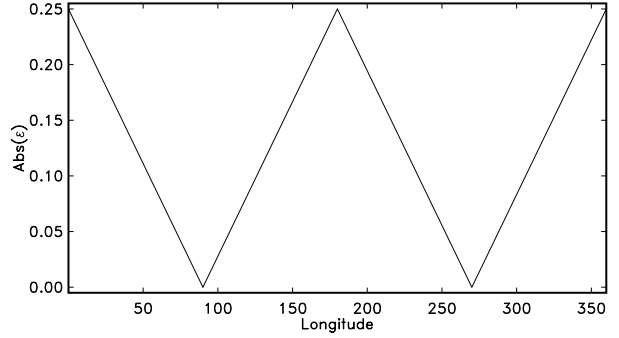}}
  \caption{The merit function depends on the distance of 
           a particular phase from the nearest 
           centre ($90\degr$ or $270\degr$). It is expected that the 
           distribution sought for will have two maxima, 
           situated $180\degr$ from each other.}
  \label{Figure06}
\end{figure}

In the actual computations described below we used 
a phase-shifted version of the merit
function, so that the expected maxima in the longitude 
distribution were shifted to
$90\degr$ and $270\degr$. 
This is to ensure uniformity of all the plots.

Thus our absolute phase values differ from UBP, however estimates of
bimodality and departure from axisymmetry are clearly unaffected
by this shift.
In the plots below we depict the merit function 
values without normalization, this does not affect the
  interpretation of the plots.
In the table~\ref{table1} we present the true (normalized)
values of $\cal E$.

The dependence of $\epsilon_{ik}$-s on the corrected phases is 
depicted in Fig.~\ref{Figure06}. 
Values of $A_{ik}$ in Eq.~\ref{equation3} are defined as
\begin{equation}
A_{ki}=S_{ki}/\sum_j{S_{ji}},
\label{equation4}
\end{equation}
where $S$ is the observed spot area corrected for the projection
effect, and the sum is taken over all the accounted for spots in the given 
Carrington rotation.

\begin{figure}
  \resizebox{\hsize}{!}{\includegraphics{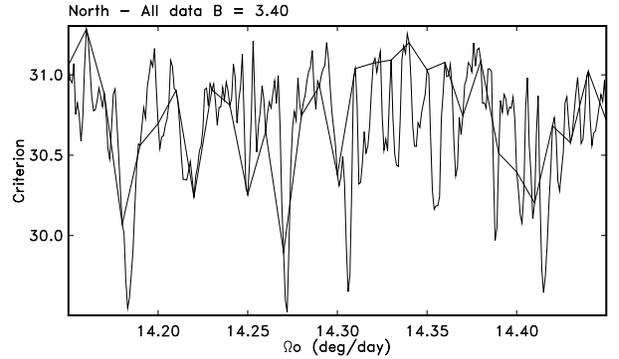}}
  \caption{The merit function dependence on $\Omega_0$ is computed with 
   two different time steps: $0.01$ and $0.001$. }
\label{Figure07}
\end{figure}
For each particular triple $(\Lambda_0,\Omega_0\,B)$ we can compute
the corresponding shifts and then evaluate the obtained phase (longitude)
distributions using the merit function ${\cal E}$. Before the actual 
computations for
a full parameter space it is important to estimate the required 
step lengths along each parameter. 

We start from the parameter $\Omega_0$. 
If we look how the shifted
phases depend on changes in this parameter, we see that a small incremental change in
the parameter, $\Delta\Omega_0$, is multiplied by the 
constant $T_C=25.38$ and then,
depending on the particular rotation, by the number of accumulated rotations. 
For the last phase the
number of rotations is approximately $1600$.

To recover full details of the merit function we fix 
a certain
maximum allowed phase shift for one step change in the input 
parameter $\Omega_0$.
Because there are two increasing and two decreasing parts 
in the distance function
(see Fig.~\ref{Figure06}), we choose one eighth of the full phase range 
($45\degr$) as the maximum
allowed phase shift. 
We obtain
\begin{equation}
\Delta\Omega_0 = 45/(T_C*1600)\approx 0.001. 
\label{equation5}
\end{equation} 
The suitability of this choice is clearly illustrated 
in Fig.~\ref{Figure07} where we
plot the merit function running along $\Omega_0$ with two different
steplengths: $0.01$ (thick curve, the stated precision in UBP) and
$0.001$ (thin
curve). Some important features are certainly lost for 
the undersampled case. (In
this plot and in all the following $\Omega_0$ or $B$ dependent plots, the
dependency on $\Lambda_0$ has been ``minimized out'' 
using the steplength $5\degr$.)
For local refinements and for short parameter intervals we have used 
even smaller values of $\Delta\Omega_0$. 
\begin{figure}
  \resizebox{\hsize}{!}{\includegraphics{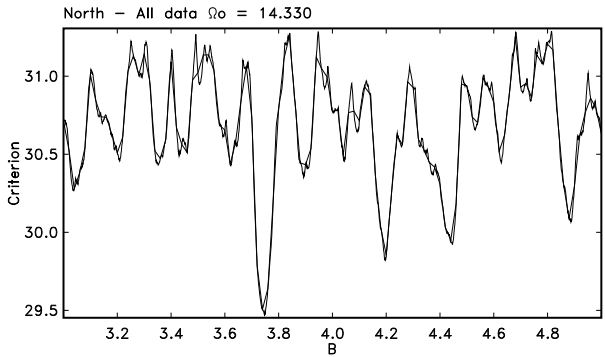}}
  \caption{ The merit function dependence on $B$ is computed with two 
different time steps: $0.02$ and $0.001$.}
  \label{Figure08}
\end{figure}

The precise analysis for the B parameter is complicated because of the 
random component in the mean latitudes. Some simple conclusions can 
be made by using
a trial and error analysis. As seen from Fig.~\ref{Figure08} the 
choice of the step
length $0.02$ (thick curve, stated precision of $B$ in UBP) is 
reasonably good,
compared with the steplength $0.001$ (thin curve). 
Nevertheless, some minima can be
slightly misplaced due to the undersampling. 
As a compromise, 
a proper
choice for the $B$ stepsize could be $0.002$.

The dependence of the merit function on the parameter $\Lambda_0$ 
is obviously smooth and we can use 
a coarser step in degrees,
say $5\degr$ to $10\degr$. Because the distance function is periodic with a
period $180\degr$ we can restrict our search to the subinterval $0\degr -
180\degr$.

We can now compute the total 
number of
trial triples we need to work with. Say we want to seek for a global merit
function minimum in the parameter space $[10-20]\times[0-6]\times[0-180]$
(according to one of the authors - I. Usoskin~(\cite{UsoskinP}), this was
exactly the chosen parameter range in their initial search). 
With the chosen steplengths, we need to evaluate at least
$(10*1000)*(6*500)*18 = 540 000 000$ triples. This is far beyond our
computational capabilities. Computations with significantly longer 
steplengths for the parameters would leave us uncertain of recovering 
the actual global minima.
\begin{figure}
  \resizebox{\hsize}{!}{\includegraphics{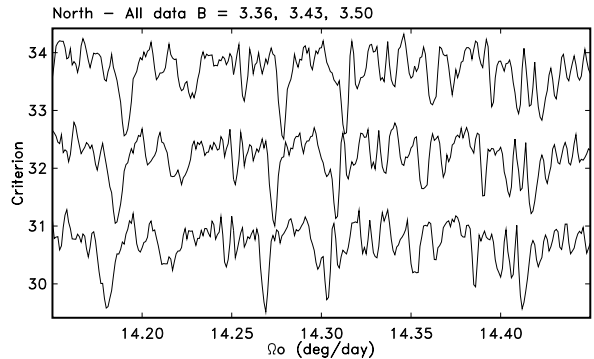}}
  \caption{The merit functions for three different values of $B$. It is 
           clearly seen that all three are quite similar 
           and only slightly shifted along $\Omega_0$. On a 
           two-dimensional diagram this would look like a row
           of slanted stripes. }
  \label{Figure09}
\end{figure}
However, we are able to reduce the search space by observing that B and $\Omega_{0}$
are not completely independent parameters.
Observing the similar shapes of curves for different values of $B$ in  Fig.~\ref{Figure09}
 and noting the structure of Eqs.~\ref{equation1.1} and \ref{equation1.2}, 
we see that
the cumulative shifts depend monotonically on the 
positive-definite sums
\begin{equation} 
S_j=\sum_{j=1}^i \sin^2 \langle \psi \rangle_j. 
\label{equation6}
\end{equation}
This component effectively consists of 
two parts: a wave which models the differential rotation during 
the solar cycle and
a linear part which depends monotonically on the sum of previous waves.
In the final
cumulative sums the fixed part of this sum is combined with a similar linear 
component which
comes from the parameter $\Omega_0$. Because these contributions are 
included in
the cumulative sums with opposite sign, we can compensate 
for the increase in the parameter
$B$ with the increase in the parameter $\Omega_0$. 
Their total contribution, which is
the difference, then stays fixed. 
The only change is a 
small redistribution in phases which comes from the cyclical part.

\begin{figure}
  \resizebox{\hsize}{!}{\includegraphics{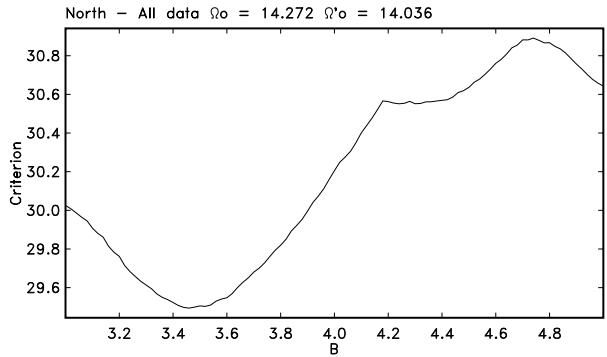}}
   \caption{For the reparameterized case the dependence on 
            $B$ is much smoother (compare with 
            Fig.~\ref{Figure08}).}
  \label{Figure10}
\end{figure}

We can now reparameterize our analysis. First we define a linear part 
of the curve
$S_j,j=1,\dots,N$. Then we subtract this linear part from the 
curve and add it to
the $\Omega_0$ dependent part. Finally we will have new parameter $\Omega_0'$
which will depend on original $\Omega_0$ and $B$:
\begin{equation}
\Omega_0'=\Omega_0-A_C\times B,
\label{equation7}
\end{equation} 
where $A_C=(S_N-S_1)/(N-1)$ is the linear slope of the 
cumulative sum of $S_j$-s.
The new parameter $\Omega_0'$ will now describe an overall constant rotation
incrementing in phase and the parameter $B$ measures the contribution from the
differential rotation.
The parameter $\Omega_0'$ now describes {\em the mean comoving 
framespeed}
and the parameter $B$ small phase shifts due to the differential rotation.The
effect of reparameterization is clearly seen in Fig.~\ref{Figure10}, the dependence
on the parameter $B$ is much smoother and it is possible to
decrease the required number of parameter triples.
We can now formulate our optimization strategy.
First we perform a coarse search using the reparameterized 
triples. The proper
steplengths for this case are $0.001$ for $\Omega_0'$, $0.1$ for 
$B$ and $5\degr$ for
$\Lambda_0$. The standard $\Omega_0$ can then be computed 
from $\Omega_0'$.
In the vicinity of the detected minima we can perform high
resolution refinements using the normal triples of parameters
with steplengths $0.0001$ for $\Omega_0$, $0.002$ for $B$ and 
$1\degr$ for $\Lambda_0$.

To characterize the search
results and estimate their significance we adopt the measure of
nonaxisymmetry defined by UBP. For each distribution of longitudes we form two
subsums of corresponding spot areas
\begin{eqnarray}
N_1&=&\sum_{k,i}{A_{ki}\,\mbox{, if }|\tilde\lambda_{ki}-90\degr|<45\degr\mbox{ or
}|\tilde\lambda_{ki}-270\degr|<45\degr},\nonumber\\
N_2&=&\sum_{k,i}{A_{ki}}\,\mbox{, otherwise,}\hskip 3.7cm
\end{eqnarray}
where the summation is taken over all the spots in all the Carrington rotations.
The nonaxisymmetry $\Gamma$ is then defined in UBP as 
\begin{equation}
\Gamma=\frac{N_1-N_2}{N_1+N_2}.
\label{Eq:gam}
\end{equation}
Non-uniform (in longitude) distributions give higher values of $\Gamma$.

\subsection{Search results}
We restricted our search space for
$\Omega_0$ to the interval $[13.5-15.0]$ since this 
includes all the
physically plausible values for this parameter. We present our results
in three parts. First we give results for fixed 
angular velocity frames, then for
frames with particular values of $B$ and finally for 
the full parameter space investigated. This
allows us to illustrate some important aspects of 
the merit function ``surface''.

\subsubsection{Fixed angular velocity frames}
From the previous discussion we know that it is quite easy to find
particular comoving frames for separate latitude strips and even mean
frames for the full data. 
Therefore we first evaluate frames with $B=0$. From UBP we
learn that``...the hypothesis of rotation of the active longitudes
with a fixed rate (giving $\Gamma = 0.02-0.03$; see Figs. 2a, 3a, and
7a,b) cannot be distinguished from the null hypothesis of the
axisymmetric sunspot distribution''. Our results of the subset search
with $B=0$ are depicted in Fig.~\ref{Figure11} and in table~\ref{table1}.
\begin{figure}
  \resizebox{\hsize}{!}{\includegraphics{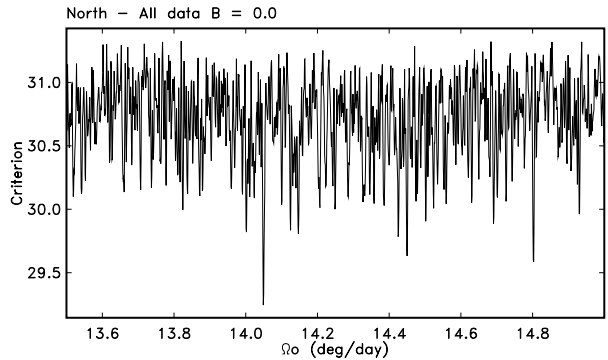}}
  \caption{The merit function computed with $B=0$  for the 
           Northern hemisphere.}
  \label{Figure11}
\end{figure}
Comparing Fig.~\ref{Figure11} with Fig.~\ref{Figure01}, we observe a key
difference: instead of the wide depression we see a number of sharp minima.
The strongest peaks are located at $\approx 14.05$ for 
North and $\approx 13.85$ for
South. The nonaxisymmetry measure is  $\Gamma = 0.064$ (North) and
$\Gamma = 0.073$ (South), instead of $0.02-0.03$ in the original paper.
We are dealing here (and also below) with the fluctuations themselves, therefore
error bars (e.g. about some mean value) are not relevant.
\subsubsection{Frames with fixed values of $B$}
To make straightforward comparisons with the UBP results we 
sought merit
function minima of slices with fixed $B$ values. For 
the Northern hemisphere we
tried to recover the original result in a slice with 
the best UBP value of $B=3.40$; 
for the Southern hemisphere our slice was selected with the best 
UBP value of $B=3.39$. The results are given in
table~\ref{table1}. 

\begin{table*}
\centering
\begin{tabular}{|lc|cc|cccc|ccc|} \hline 
Comment &N/S &$B^{comov}$ &$\Omega^{comov}_0$ &$B^{contra}$ &$\Omega^{contra}_0$ &$\cal E$ & $\Gamma$ & $B^{ubp}$ & $\Omega^{ubp}_0$ & $\Gamma^{ubp}$ \\ \hline
Fixed B &N &-3.40 &14.0971 & 3.40 &14.2717 &0.0195 &0.059 &3.40 &14.33 &0.11 \\ 
Fixed B &N & 3.40 &14.3652 &-3.40 &14.0036 &0.0195 &0.064 &-       &-         &-       \\ \hline
Fixed B &S &-3.39 &13.6344 & 3.39 &14.7344 &0.0196 &0.038 &3.39 &14.31 &0.09 \\ 
Fixed B &S & 3.39 &14.6170 &-3.39 &13.7518 &0.0192 &0.070 &-       &-         &-       \\  \hline
No diff. rot       &N &0        &14.3196 &0       &14.0492 &0.0194 &0.064 &0      &?         &0.02-0.03 \\ 
Global search &N &-1.48 &14.3183 &1.48  &14.0505 &0.0191 &0.079 &3.40 &14.33 &0.11          \\
Global search &N &3.92  &14.4395 &-3.92 &13.9293 &0.0190 &0.096 &-       &-         &-                \\ \hline
No diff. rot       &S  &0       &14.5193  &0      &13.8495  &0.0194 &0.073 &0      &?         &0.02-0.03 \\ 
Global search &S  &-0.63 &14.4992 &0.63  &13.8696 &0.0191 &0.068 &3.39 &14.31 &0.09          \\
Global search &S  &3.72  &14.6404 &-3.72 &13.7284 &0.0190 &0.082 &-       &-         &-                \\
\hline
\end{tabular}
\caption{Summary of the search results with the UBP method. The analysis is performed in
the contramoving formulation: the obtained frame parameters are given in the 
columns 5 and 6, and transformed into the comoving formulation in columns 3 and 4. The corresponding best values from the original UBP paper, if available, are given in columns 9 and 10.} \label{table1}
\end{table*}

In addition to the computations with positive values of $B$ we computed
``spectra'' also around their negative counterparts. 
It is clearly seen from the corresponding
tables that our results differ significantly from these presented in UBP.
\begin{figure}
  \resizebox{\hsize}{!}{\includegraphics{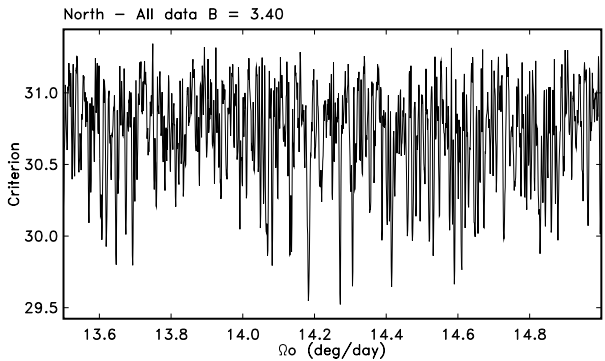}}
  \caption{The merit function computed with $B=3.40$ for the Northern 
   hemisphere. This figure should be  compared with Fig.~\ref{Figure03}.} 
  \label{Figure12}
\end{figure}
Fig.~\ref{Figure12} is of particular significance when compared with
Fig.~\ref{Figure03}, where we can see that the best comoving frame shows itself as 
a very clear minimum in the middle whose width is much greater than that of
the fluctuations. 
In Fig.~\ref{Figure12} we observe  only a bunch of narrow fluctuations.
With the merit function used in section~\ref{Section:cellcounting},
when we change the parameter values, then in the vicinity of 
the main minimum, phases migrate
slowly from one cell to another and the general 
picture changes slowly. With the UBP merit function
(crafted for bimodal waves), 
we observe a highly fluctuating curve, with no observable minimum of width
greater than the fluctuations.

\subsubsection{Global search}
In the global search we allow $B$ also to vary freely. 
The results of the global computations are summarized 
in table~\ref{table1} and in Fig.~\ref{Figure13}.
The distribution for the Northern hemisphere does not show the clear
  bimodal structure presented in UBP, but its nonaxisymmetry is quite high -
  $\Gamma = 0.096$.
Our results with the contramoving formulation and UBP merit function
show that there are no minima with width greater than the fluctuations which
would be expected from causal dependence on the parameters.
Given this, we observe that the strongest minima tend to occur for 
the negative values of $B$,
which translate to solutions with the comoving frame formulation. 
\begin{figure}
  \resizebox{\hsize}{!}{\includegraphics{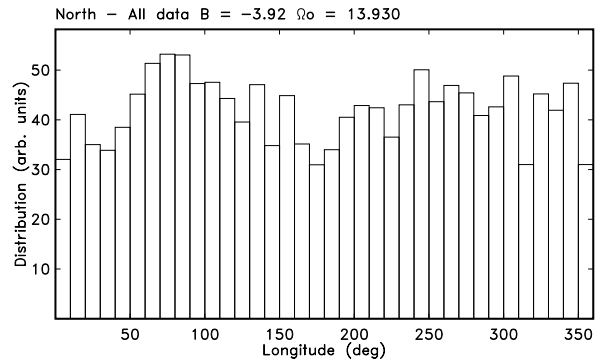}}
  \caption{Strongest fluctuation for Northern hemisphere.}
  \label{Figure13}
\end{figure}
\begin{figure}
  \resizebox{\hsize}{!}{\includegraphics{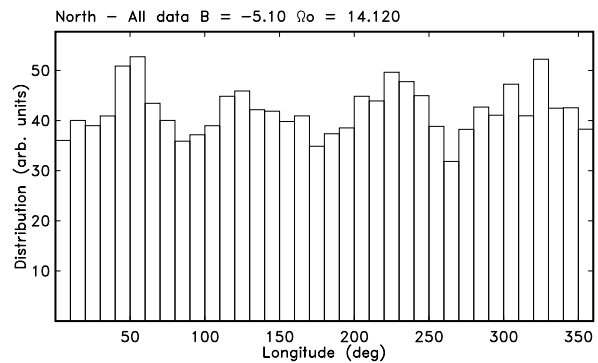}}
\caption{Longitudinal distribution obtained with best parameter fit for a
  merit function with four maxima.
  The parameter space used in the search was the same as in Fig.~\ref{Figure13}.}
  \label{Figure14}
\end{figure}

Possible reasons for the difference between our results and
the results presented in BU and UBP can be:
\begin{itemize}
\item The data sets used in UBP and here can be different. The authors of UBP may have used certain pre- or
  postselection principles applied to the input data (not described in the paper).
\item The curves of the mean latitudes (and therefore also the shift frames) in UBP were
  smoothed in a particular way (see Fig. 1 in UBP) or were interpolated in
  a particular way (compare Fig. 5 in BMSU to our Fig.~\ref{Figure15}). However, we did not
  find any comments about these procedures in either of the papers.
\end{itemize}
Below we try to look at some other possibilities. 
To check if the bimodal merit function gives clearly superior results for
departures from axisymmetry, we constructed a merit function with four maxima
over the full range of longitude and performed a full parameter space search as above.
This represents not two migrating density waves but four, two for each side of the Sun. 
The fluctuation
which gave the 
global minimum for this criterion is depicted 
in Fig.~\ref{Figure14}
(Northern hemisphere).
The distribution is certainly non-uniform and four expected waves are clearly seen.

\section{Discussion}\label{Section:discussion}
The results obtained above and those of BU, UBP and BMSU (data analysis part)
show discrepancies which require some explanation. Below we present how we
tried to control our procedures to discover potential reasons for such major
differences.
We also include comparative analysis of the results of BU and UBP since a
major contention of PTB is that the attempt to detect persistent features of the sunspot data is very
sensitive to the methodology used.  

\subsection{Kinematics of frames}
The most significant problem with the persistent longitude strips is connected to 
their cumulative kinematics. 

The authors of BU found that in the Carrington reference frame the
observed migration had a phase lag about 2.5 solar rotations per
sunspot cycle, in total about 28 rotations during 120 years.

In the follow-up study (PTB) we demonstrated that a major part of this
effect can be due to hidden assumptions in the methodology of data processing.
In the next paper (UBP) the authors attempt to confirm their original
conclusions by
constructing a new set of contramoving frames using global
optimization techniques. The selected dynamic frames, however, differ
significantly between BU and UBP \& BMSU: in the former case the
pattern makes 28 extra rotations during 120 years, whereas in the
latter, the Northern pattern makes $\approx 11$ and Southern $\approx
14$ rotations. The reader can easily verify this by comparing
Figs.~3-4 from BU with Fig. 1 in UBP and especially with Fig. 1 in
BMSU, where the different frames and their kinematics are
depicted graphically. Consequently we consider that the two latest papers (UBP and
BMSU) fail to confirm the results obtained in the first paper (BU), in
particular they do not refute the conclusions of PTB.

The second kinematical problem is connected to the different parameterization
schemes. 
In the UBP investigation, a contramoving frame system is used, in
addition restricted to positive values of the parameter $B$. As
discussed in length in Sect.~\ref{Subsection:comoving}, to construct a frame system
comoving with the solar surface differential rotation pattern,
negative $B$ values should have been allowed for. Now the solutions
are restricted to trace a situation where movement in the opposite
direction than the differential rotation pattern occurs.
Since we have demonstrated in section~\ref{Subsection:comoving} that these frames are
not equivalent methods of treating the observational data, 
the restriction of positivity of $B$ should be
dropped, as in this study. As a consequence, with the UBP method,
we find the highest nonaxisymmetry measures for frames with negative
$B$s, i.e. moving with the solar surface differential rotation
pattern.

The third kinematical problem comes from the fact that the numbers of extra rotations
(compared with Carrington frame) are in the UBP model significantly different for
the Northern and the Southern hemispheres: $\approx 11$ and $\approx 14$
correspondingly (cf. Fig.~1 in BMSU). This raises serious difficulties for
the dynamo theory of the large scale magnetic field.

\subsection{Evidence of flip-flops in the dominant longitude}
Another important result of BU is the alternation of dominance 
between the two longitudinal maxima  
with the approximate periods of 3.8 (North) and 3.65 (South) 
years (periodicity in flip-flops). The flip-flop phenomenon was 
originally found in an extremely active late-type star FK Com
(see Jetsu, Pelt \& Tuominen~\cite{Jetsu93}).

The particular values for the mean flip-flop periods for the Sun (in BU) 
were obtained 
from an analysis of the power spectra. 
Firstly the evidence presented does not establish that such periodicity exists. 
From the power spectrum plot of Fig.~8-9 (BU) it is clear that 
{\em nearly all the spectral power 
fluctuations} are above the line of $95\%$ confidence level.
If there were clear and dominant periodicity one would not expect this.

In UBP it is claimed that the flip-flop events and their periods obtained in BU are valid 
also for the newly constructed dynamic frames and the resulting coherent
nonaxisymmetric structures. 
The flip-flop event 
frequency, however, was computed using a frame which rotates
nearly three times more rapidly than the dynamic frames found in UBP.
We note that a phase shift
of $\approx 180\degr$ which causes the flip-flop event position to swap longitude can
be produced by $0.004$ degrees per day change in the frame model parameter
$\Omega_0$ (quoted precision in UBP for this parameter is $0.01$). In the new
construction, in addition, the shifts are cumulative sums of functions
from a significantly fluctuating mean latitude curve (see
Fig.~\ref{Figure15}). Thus evidence for the persistence of this phenomenon across
the two methods of determining rotating frames is not sufficiently established.
\begin{figure}
  \resizebox{\hsize}{!}{\includegraphics{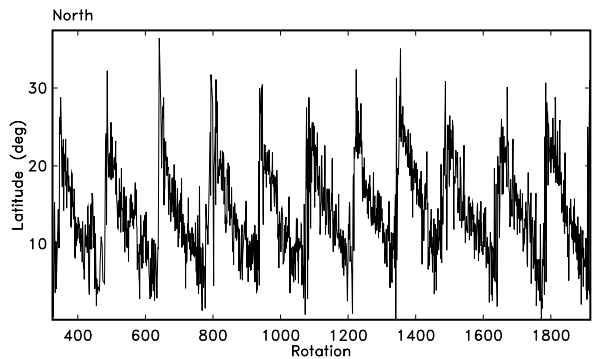}}
  \caption{The mean latitude of sunspots(weighted by sunspot area) against
    time. The magnitude of the fluctations against the general trend can be seen.} 
\label{Figure15}
\end{figure}

Consequently it is not reasonable to combine the results obtained from one frame
system with the results of a totally different system.

\subsection{Nonaxisymmetric distributions in fixed velocity frames}
It is a major point of the UBP analysis that measures of nonaxisymmetry are
low ($\Gamma = 0.02-0.03$) for frames with fixed rotation (see
Fig.~7 in UBP). In our
computations we obtained a relatively high $\Gamma$ for fixed rotation frames (see
Table~\ref{table1}), of the same order as the best values for differentially
rotating frames. Since we have demonstrated that the UBP merit function shows
rapid fluctuation with respect to changes in the parameter values, it is 
also very
possible that results are sensitive to preprocessing of data. Accordingly we
have published on the web the exact data that we used
(c.f. section~\ref{Subsection:comoving}). 
We think that in
looking for evidence of persistent longitudes it is necessary to examine
parameter space very carefully and be completely explicit about all data
processing steps used, with publication of data as an aid to other researchers.

\subsection{The effects of folding phase shifts}

In the process of seeking other possible reasons for differences between our computations and those of UBP we folded 
(by $360\degr$) the shift curves with different parameter values.  
\begin{figure}
  \resizebox{\hsize}{!}{\includegraphics{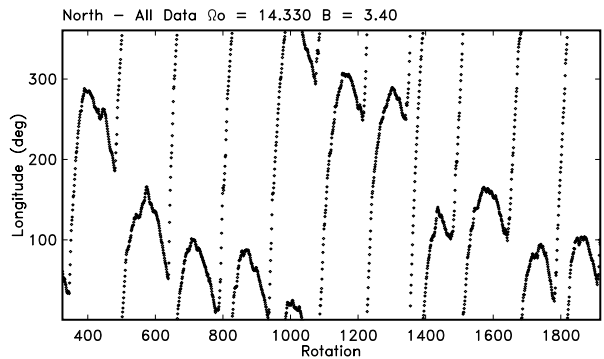}}
  \caption{Particular shifts for a contramoving frame with 
$\Omega_0=14.330$ and $B=3.40$ folded with module 
   $360\degr$. }
  \label{Figure16}
\end{figure}
\begin{figure}
  \resizebox{\hsize}{!}{\includegraphics{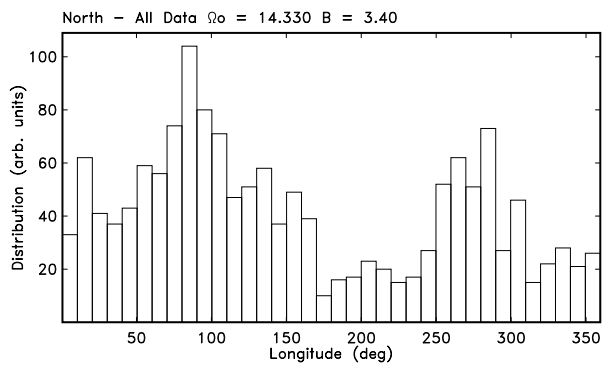}}
  \caption{The marginal distribution (along longitudes) of the frame depicted in Fig.~\ref{Figure16}.}
  \label{Figure17}
\end{figure}
In Fig.~\ref{Figure16} the shift curve for the UBP contramoving frame is
depicted.
The key features in this plot are the ``S''-shaped fragments. 
For different frame parameters these fragments line up differently.
The corresponding marginal distributions  in Fig.~\ref{Figure17} show that the
shifts themselves are not distributed evenly. 
In Fig.~\ref{Figure18} we have depicted how the
merit function changes for a fixed value of $B=3.40$ and changing value of $\Omega_0$. 
\begin{figure}
  \resizebox{\hsize}{!}{\includegraphics{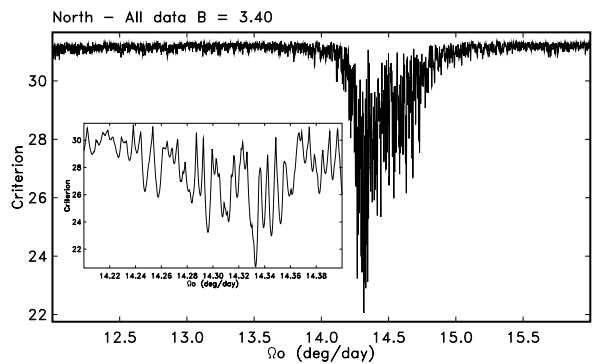}}
  \caption{The merit function computed with longitudes set to zero. There is 
   a strong depression 
   around $\Omega_0=14.20-14.40$. Enlargement of the depression part is given as inlet. }
  \label{Figure18}
\end{figure}
The plot demonstrates
that the merit function has a specific wide depression; in the bottom of this
depression around $\Omega_0$ values $14.20-14.40$ there are deeper minima.
Fig.~\ref{Figure18} (inlet) reveals that the deepest minimum occurs at $\Omega_0=14.316$. The second deepest is at
$\Omega_0=14.330$. 
Comparison of  
Fig.~\ref{Figure17} and 
Fig.~\ref{Figure19} shows that the value $\Omega_0=14.330$ is {\em the strongest peak
whose shift system has a characteristic double mode distribution}. 
Just to remind the reader, we
{\em did not use longitudes at all} to compute the last three plots.
We can see how sensitive the form of the distribution is to parameter
values giving merit function minima of comparable strength.
\begin{figure}
  \resizebox{\hsize}{!}{\includegraphics{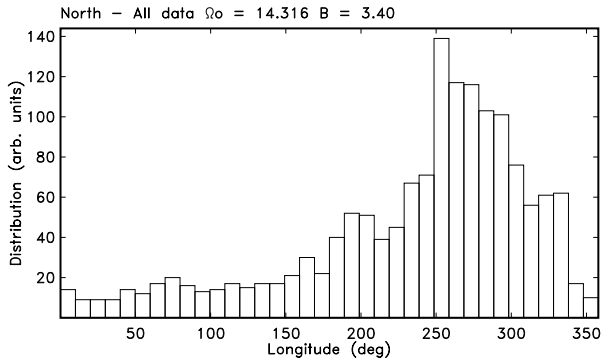}}
  \caption{The marginal distribution for the frame with $\Omega_0=14.316$ and
    $B=3.40$.  (compare with Fig.~\ref{Figure17}).}
  \label{Figure19}
\end{figure}

We draw also attention to the particular value of $\Omega_0=14.33$ deg/day, which occurs both as a solution in UBP 
and as fluctuations in the input data (shift curve) for the Northern and Southern hemispheres: is there 
anything peculiar behind this coincidence?
The unexpected solution can be found with the help of the previous analysis. 
To compute minima for the large parameter grid we used certain reparameterization where we
divided the frame shift curves into two parts: the linearly rising part and 
the cyclical part.
From the linear part we computed then mean rotation velocity for the frame. For
the particular values $B = 3.40$ and $\Omega_0 = 14.330$ this gives $\Omega_0' =
14.094$. The charming magic of this value reveals itself when we compute
first how much it differs from the Carrington siderial value 
$\Omega_C-\Omega_0' =
0.090$. From the angular velocity we can compute the corresponding period which occurs to be 
$10.96$ years - closely coinciding with the mean solar 
sunspot cycle length. The obtained result can also be 
read out from Fig.1 in BMSU - there are approximately 11 cycles and the range of the shifts is also approximately 
11 full rotations.

\subsection{Selection of a correct null hypothesis}

We now consider the selection of a null hypothesis to check the significance
of evidence for persistence in the longitudes of sunspot activity. 
In UBP the significance of
the obtained results was evaluated using a reference (null hypothesis) distribution which was computed
using Monte-Carlo type methods. For each Carrington rotation they randomly
permuted all the sunspots, i.e., {\em a new set of random Carrington longitudes} were
ascribed to each actually observed sunspot while keeping its area. Then the
value of $\Gamma$ was calculated as described above. They computed the
nonaxisymmetry $\Gamma$ for 5000 sets of such random-phase sunspot occurrences
to get a reference distribution. This is a valid procedure for the
case when we need to test certain particular longitude distributions against
totally random distributions. 

However, this analysis does not take into account the fact that we select
  the best fit
solution among number of candidate solutions. Consequently we need to compute the 
``false alarm probabilities'' (see for detailed explanation
Scargle~\cite{Scargle}). Thus in the Monte-Carlo
framework we must compute distributions for ``record values''. 
The full scale, full resolution Monte-Carlo computation for
full $\Omega_0$,$B$ parameter space is certainly out of our computational
capabilities. 
Fortunately, the general picture becomes evident from a subset of calculations.
\begin{figure}
  \resizebox{\hsize}{!}{\includegraphics{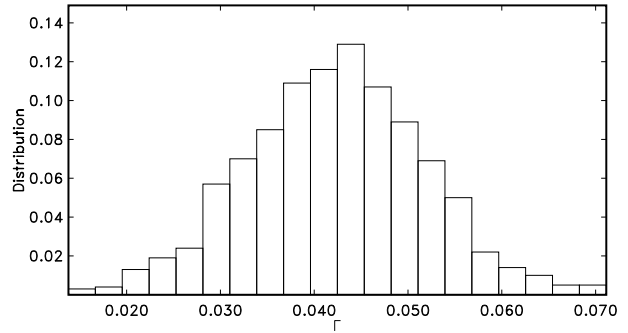}}
  \caption{Distribution of records for fully decorrelated data.}
\label{Figure20}
\end{figure}
\begin{figure}
  \resizebox{\hsize}{!}{\includegraphics{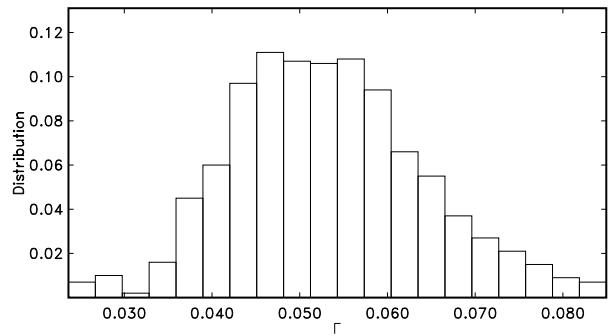}}
  \caption{Distribution of records for short-term correlated data.}
\label{Figure21}
\end{figure}
 First we restricted our computations to parameter space slices with
  fixed parameter $B=3.40$ and a grid with moderate resolution for $\Omega_0$ -
  [13.5,15.0,0.01] (Northern hemisphere). Then we performed for 1000
  randomized (as in UBP) data sets search for strongest peak in $\Omega_0$
  spectrum. For the strongest peak we computed the value for $\Gamma$.
  The corresponding value distribution is depicted in Fig.~\ref{Figure20}. It is
  evident that, even for fully decorrelated data, the record values can be
  quite high. This can be explained by the fact that all data points are not
  equal - they are weighted by area. The areas differ very significantly (from
  0 to 1894 millionths of total area) and the number of effective degrees of
  freedom to be distributed along longitudes is significantly less than total
  number of activity elements (18680).
    
It is known that in the Sun we can
  observe quite long-lived activity complexes, for instance hot spot systems
  of solar flares with lifetimes of the order of the solar cycle length
  (Bai~\cite{Bai88,Bai03}).  To model this kind of situation we performed
  an additional test. Instead of full randomization of the longitudes we
  divided
  the full observational data into 10 rotation length sections. For each of the
  1000 Monte-Carlo runs we added random shifts from the interval $[0,360]$ to
  every section. In this way local correlations remained untouched, but long
  range correlations (if present) were certainly washed out.  The record
  distribution for this experiment is depicted in Fig.~\ref{Figure21} and it is
  to be compared with results in table~\ref{table1}. We see that we can safely
  treat global minima as the random fluctuations due to the local correlations
  and the uneven distribution of sunspot areas.
  Our analysis is conservative - we undersampled along $\Omega_0$ parameter,
  we
  used a fixed value of $B$ and we did not seek minima in $\Gamma$ but in the UBP
  merit function.  

In addition to the above mentioned correlations there is another source of flexibility in
the UBP method which results in additional fluctuations. 
This is due to the reweighting procedure chosen. 
In the input data set there are some rotations which
contain nearly 70 new activity elements and all these get minuscule weights in
the method. 
The weights for the elements which happen during the rotations between
the activity cycles, on the other hand, are very powerfully amplified because 
these rotations are
strongly underpopulated (rotations with only a single new event occur
reasonably often). As a result, many kinds
of distribution may form (see for instance Fig.~\ref{Figure14}).  Instead of
amplifying the effect of persistent activity waves 
the method amplifies random fluctuations.

\subsection{Inherent constraints of the data set.}

Given the difficulties we have described above we examine the constraints
caused by the limited time span of the data set, relative to the solar cycle.
The normal comoving flow of
the activity indicators can be appropriately measured by such methods. But for activity
waves which travel in the opposite direction to solar rotation the evidence can 
only be quite a weak difference between a low level real physical effect and 
the fluctuations which result from the statistical nature of the data at hand. 

As shown in PTB it is possible to combine a small number of long stretches of
data (using appropriate phase corrections) so that their marginal distribution
is almost entirely an artefact of the method of analysis.
 
The inherent (for such an analysis) phase ambiguity is a strong constraint
which can be overcome only by using much longer (in time) data sets.
The current time extent of the sunspot database covers only a small number of correlation
length size subparts and this results in a high level of random fluctuations. We
do not deal here with 1600 independent rotations but with a significantly lower
amount of activity complexes.

\section{Conclusions}\label{Section:conclusions}
In the recent papers BU, UBP and BMSU the authors claim that, hidden in the sunspot
distribution, there are two persistent active longitudes which
migrate according to a certain differential rotation law. After presenting a critique of the
BU results in the earlier paper (PTB), here we have tried to
check the results obtained in UBP and the data analysis part of BMSU. 
Our results can be summarized as follows:
\begin{itemize}
\item The contramoving frames constructed by BU  differ significantly from
  the frames found using mathematical optimization techniques in UBP and BMSU. In
  the first case the frame does approximately 28 extra rotations (compared
  with the Carrington frame) and in the new papers the number of extra rotations
  is around 10-11 for the Northern hemisphere and 13-14 for the Southern hemisphere.
  Consequently the claim in UBP that the new analysis confirms the results
  obtained in the previous study is not well founded.
\item In the two latest papers authors re-iterate the claim presented in BU of the 
presence of the flip-flop 
  effect with a mean period of 3.7 years. This result was obtained
  from (and now presumably incorrect) analysis in the first paper (BU).
\item {Our analysis demonstrated that even for fixed angular velocity 
  contramoving frames it is
  quite easy to get particular longitude distributions with relatively high
  values of nonaxisymmetry. This result is quite robust and does not depend
  on the particular implementation of the UBP method (smoothing of the shift
  curve, sampling rates in the parameter space etc.). Consequently all 
  the claims
  in UBP about $\Gamma = 0.02-0.03$
  level fluctuations in the case of $B=0$ are incorrect. The parameter space is 
  full of parameter combinations 
  which can give a rather high values for nonaxisymmetry.}
\item Our re-implemantion of the UBP method gave fundamentally different results from those of the original
  authors. We can only conjecture about
  the possible reasons for the discrepancies. The pre- or postselection of data points,
  parameter undersampling in search procedures, incorrect phase reduction etc.
  - these all can be considered as possible sources of differences.
\item {From inspection of the strongly fluctuating merit function curves
    and additional 
Monte-Carlo type computations for parameter space slices, we conclude that the frames obtained from
  the optimization procedure are pure fluctuations.}
\item We found that the parameter values obtained as results in UBP can be seen
  already in the distributions connected to the frame shift curve itself
  (which does not depend on longitudes of sunspots at all).

\item Our analysis does not indicate that short-lived features concentrated in certain
longitude are totally missing in the sunspot distribution, even though evidence for century 
scale persistent active longitudes can not be found. The simple cell-counting method 
investigated here merely as a demonstration tool may well be used to estimate real coherence 
times of these structures.
\item {Whichever method used, the preferred solutions occurred at positive/negative values 
of the differential rotation parameter $B$ in the comoving/contramoving frame systems,
respectively, i.e. the frames had a tendency to follow a solar-type differential rotation pattern.}    

\end{itemize}
From this we conclude that the results obtained in BU, UBP and BMSU (data analysis part) are 
inconsistent, and evidence for well established
persistent activity migration is still lacking. In other words we are inclined to think 
that the enigmatic face of the Sun is
still hidden and that the patterns which the authors of the three papers saw have 
a human origin.

\begin{acknowledgements}
We are thankful to anonymous referee and A. Brandenburg for helpful comments on the manuscript and I.G. Usoskin for additional
comments about data processing procedures used in the UBP paper.
Part of this work was supported by the Estonian Science Foundation
grant No. 6813 and Academy of Finland grant No. 1112020. 
\end{acknowledgements}

\end{document}